\documentclass[manuscript]{aastex61}

\received{2019 April 9}
\accepted{2019 July 22}
\published{2019 September 20}
\shorttitle{Numerical simulation of magnetic reconnection around black hole}
\shortauthors{Inda-Koide, Koide, and Morino}
\begin{document}

\title{Numerical simulation of magnetic reconnection around a black hole}

\author{Mika Inda-Koide}
\affiliation{\rm Luther Senior High School,
3-12-16, Kurokami, Chuo-ku, Kumamoto, 
%860-8520, Japan}
860-8520, Japan; koidesin@kumamoto-u.ac.jp}

\author{Shinji Koide}
\affiliation{\rm Department of Physics, Faculty of Science, Kumamoto University,
2-39-1, Kurokami, Chuo-ku, Kumamoto, 860-8555, Japan}

\author{Ryogo Morino}
\affiliation{\rm RKK Computer Service,
1-5-11, Kuhonji, Chuo-ku, Kumamoto, 862-0976, Japan}



\begin{abstract}
We performed numerical simulations of general relativistic 
magnetohydrodynamics with uniform resistivity to investigate
the occurrence of magnetic reconnection in a split-monopole magnetic
field around a Schwarzschild black hole. 
We found that magnetic reconnection
happens near the black hole at its equatorial plane.
The magnetic reconnection has a point-like reconnection region and
slow shock waves, as in the Petschek reconnection model.
The magnetic reconnection rate decreases as the resistivity becomes smaller.
When the global magnetic Reynolds number is
$10^4$ or larger,    
the magnetic reconnection rate
increases linearly with time   
from $2 \tau_{\rm S}$
to $\sim 10 \tau_{\rm S}$ ($\tau_{\rm S}=r_{\rm S}/c,
r_{\rm S}$ is the Schwarzschild 
radius and {\it c} is the speed of light).
The linear increase of  
the reconnection rate
%as a function of time or resistivity 
agrees with the 
magnetic reconnection in the 
Rutherford regime of the tearing mode instability.  
\end{abstract}




\section{Introduction} 
Galaxies classified as a class of active galactic nuclei (AGNs)
are believed to
harbor supermassive black holes in their centers. \rm Stellar mass black holes
are also thought
to reside in long-duration $\gamma$-ray bursts (LGRBs) and 
microquasars in our Galaxy. 
From some of these black holes, plasma ejections, like moving  
radio knots or jets, are
observed. As an example of radio knot ejections from AGNs,  
%the VERITAS Collaboration et al. (2009)
Acciari et al. (2009)
%\citet{acciari09}
presented simultaneous radio and $\gamma$-ray observations of the nearby active galaxy
M87, and showed that radio knots were ejected from the core of the galaxy when
the TeV $\gamma$-ray flare occurred. As for relativistic jets from AGNs,  
Biretta et al. (1999)
reported observations of superluminal motion in the M87
jet by the \it Hubble Space Telescope. \rm    
Kulkarni et al. (1999) presented optical and near-infrared observations
of the afterglow of GRB990123, 
and argued that the detected $\gamma$-ray is
relativistically beamed. 
%until the early phase of the afterglow. 
Mirabel \& Rodriguez (1994) reported superluminal motion 
of the radio-emitting ejecta
from the center of \rm  
microquasar GRS 1915$+$105.
These plasma boosts are believed to be  
caused by violent phenomena around black holes. 

Such plasma ejections are also observed in solar flares (e.g., 
Ohyama \& Shibata 1998). 
Solar flares and associated coronal mass ejections are phenomena
related to solar
magnetic field: they exhibit a sudden energy release of magnetic energy, 
which is
transformed to  electromagnetic radiation energy (flares) or mechanical
kinetic energy (coronal mass ejections). 
An idea called magnetic reconnection was proposed to explain 
solar flares in the 1940s (Giovanelli 1946), and in 
the 1950--1960s the 
magnetohydrodynamics (MHD) theory of magnetic reconnection was constructed 
by Sweet (1958), Parker (1957), and Petschek (1964), known as the 
Sweet--Parker and Petschek magnetic reconnection models.
An important physical quantity of magnetic reconnection is the magnetic 
reconnection rate, $R_{\rm mr}$. 
$R_{\rm mr}$ expresses the time change rate of magnetic flux at the 
reconnection point. From the Faraday's law, the 
parallel component of 
the electric field to the reconnection line (X-line) 
at the reconnection point, $E_\parallel$, represents this rate.
By normalizing $E_\parallel$ with the physical quantities of the plasma, 
$R_{\rm mr}$ is obtained. 
From a theoretical point of view, $R_{\rm mr}$ 
depends on the reconnection type and stage.
The Sweet--Parker and Petschek
models describe the stationary (steady) state of magnetic 
reconnection, so $R_{\rm mr}$ is time-independent in these models.
As for the resistivity-dependence of 
$R_{\rm mr}$, $R_{\rm mr} = 1/\sqrt{S_{\rm e}}$ 
for the Sweet--Parker model, and
$R_{\rm mr}$(max) $= \pi / (8 {\rm ln} S_{\rm e})$
for the Petschek model.
Here, $S_{\rm e}$ is the global magnetic 
Reynolds number and 
%or the global Lundquist number,
$R_{\rm mr}$(max) is the maximum magnetic reconnection rate 
estimated by Petschek.
We give exact definitions of $R_{\rm mr},   
S_{\rm e}$,
and the magnetic Reynolds number $S$   
in Section 3. 
Both $S$ and $S_{\rm e}$ are inversely proportional to resistivity 
(Parker 1957; Sweet 1958; Petschek 1964; Priest \& Forbes 2000; Kulsrud 2005).
Considering the typical time scale of a flare, $10^2-10^4$ s,
the Petschek model better explains the rapid energy release in a flare
than the Sweet--Parker model (e.g., Kulsrud 2005; Shibata \& Magara 2011). 
Here we mention that for the occurrence of the Petschek-type reconnection, 
the local enhancement of resistivity seems to be an essential
process
(but also see Baty et al. 2009 for a different consideration;
we describe it in section 3)
(Shibata \& Magara 2011).

Before magnetic reconnection, or a current sheet,
settles down to the stationary stage as the
Sweet--Parker or Petschek-type reconnections,  
the sheet is subject to the tearing mode instability
(Furth et al. 1963), which is a
nonstatic state. 
This state starts from the linear growth stage
(Furth et al. 1963)
and then shifts to the nonlinear
growth stage called the 
Rutherford regime (Rutherford 1973; Murphy et al. 2008).
We illustrate the time development of magnetic reconnection ($R_{\rm mr}$ 
as a function of time) in plasma with uniform resistivity and an initially 
uniform current sheet in Figure 1.
During the tearing mode instability, $R_{\rm mr}$ varies with time.
From the theories, $R_{\rm mr}$ is proportional to  
$e^{\gamma t}$ where $\gamma \propto 
1 / \sqrt{S}$ in the linear growth stage
(Furth et al. 1963; Biskamp 1993) 
and $R_{\rm mr}$ is proportional to $t$
in the Rutherford regime (Rutherford 1973). 
The tearing mode instability was first 
proposed as the theory to disruptions in 
laboratory fusion devices such as tokamaks.
This instability eventually introduces nonuniformity into 
the current sheet where a series of magnetic islands or plasmoids is formed.
Nonsteady reconnection associated with multiple magnetic islands often causes
impulsive bursty reconnection
(Priest 1985; Shibata \& Magara 2011).
%(Shibata \& Magara 2011).
The tearing mode instability, especially in the nonlinear Rutherford regime,
is a complicated process,
but it is important to understand time evolution of 
magnetic reconnection, and also the intermittent energy release
(nonsteady reconnection) commonly observed in the Sun, terrestrial
magnetosphere, and laboratory plasmas.

\begin{figure} % [H]
\begin{center}
\includegraphics[width=9cm]{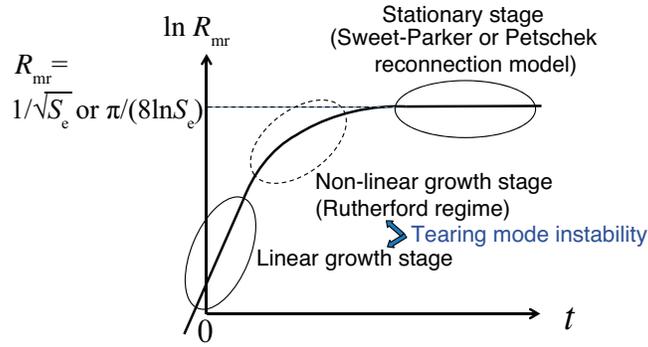}
\caption{
Schematic picture of
time development of magnetic reconnection from the early stage to
the stationary stage in plasma with
uniform resistivity and initially uniform current sheet.
$R_{\rm mr}$ is the magnetic reconnection rate, $S$ is the magnetic
Reynolds number, and $S_{\rm e}$ is the global magnetic Reynolds number.
In the linear growth stage, $R_{\rm mr} \propto e^{\gamma t}$ where $\gamma
\propto 1/\sqrt{S}$, and $R_{\rm mr} \propto t$ in the Rutherford regime.
Finally,
$R_{\rm mr}$ settles down to the stationary (time-independent) stage as
the Sweet--Parker or Petschek
type reconnection. In this stage, $R_{\rm mr} = 1/\sqrt{S_{\rm e}}$ or $ 
\pi/(8{\rm ln}S_{\rm e})$.
\label{fig1}}
\end{center}
\end{figure}

The magnetic reconnection can also be a key process to  
cause ejection from black holes.
To create an occurrence of magnetic reconnection around a black hole,
an anti-parallel magnetic field with a current sheet needs to exist there.
We here discuss the 
possibility of spontaneous formation of an antiparallel
magnetic field
around a black hole.
First, let us assume that the initial magnetic field has the uniform 
magnetic component, aligned with the rotational axis
of a rapidly spinning black hole.
In this magnetic configuration, one may think
that an antiparallel magnetic field is  
scarcely formed. 
However, even in the initially uniform magnetic field, 
the simulation of the general relativistic MHD with
zero electric resistivity (ideal GRMHD) 
showed that the magnetic flux tubes 
become stationarily   
antiparallel  
in the ergosphere around the  
equatorial plane of the black hole 
(Komissarov 2005).
Next, as the initial condition, we consider  
magnetic flux tubes
connecting an accretion-disk around a rotating black hole and a black hole
ergosphere.
With this initial condition, 
ideal GRMHD simulations also indicated that the antiparallel magnetic
field is formed spontaneously after a sufficiently long term 
(Koide et al. 2006; McKinney 2006).
%\citep{koide06,mckinney06}. 
As shown in the above two cases of magnetic field, 
an antiparallel magnetic field (the magnetic configuration where  
magnetic reconnection occurs) is relatively easily formed around a 
spinning black hole; thus,  
magnetic reconnection is expected to
happen frequently around a black hole.
This may explain observed plasma boosts from black holes.

Koide \& Arai (2008) investigated energy extraction from a rotating black hole
by magnetic reconnection in the ergosphere. As they discussed, the phenomena
of magnetic reconnection around a black hole 
should be investigated by numerical simulations of the full GRMHD with nonzero
electric resistivity (resistive GRMHD).
Two groups, to our knowledge,  
have developed numerical codes to
solve the equations of 
resistive GRMHD    
(Bucciantini \& Del Zanna 2013;  
Dionysopoulou et al. 2013).    
Dionysopoulou et al. (2013) simulated the  
gravitational collapse of a magnetized, nonrotating neutron
star to a black hole, and Dionysopoulou et al. (2015) 
studied the dynamics of binary neutron stars. 
In the work of Dionysopoulou et al. (2015), when a black hole had been formed
after the merger of two neutron stars, a magnetic-jet structure was formed
in the low-density funnel produced by the black hole--torus system; though, 
a relativistic outflow was not produced in their results.
Regarding accretion disks of AGNs, Bugli et al. (2014), with the code 
of Bucciantini \& Del Zanna (2013), considered dynamo action in thick disks 
around Kerr black holes.
However, no simulation work with resistive GRMHD 
has been released that shows clear magnetic reconnection feature in plasma
in the universe.   

In this paper, we report the results of numerical simulations using 
the resistive GRMHD code that we developed to investigate
the occurrence of magnetic reconnection in plasma (possibly an accretion
disk) just outside a 
black hole. We numerically 
explored a time evolution of physical quantities in the field.
In our results, magnetic reconnection is for the first time 
clearly visible in plasma near a
black hole.
To investigate the basic physics of magnetic reconnection around a
black hole, we assumed a Schwarzschild black hole, and 
split-monopole
magnetic field around it 
as the initial condition of the magnetic field.  
We found that relatively fast magnetic reconnection
happens near the black hole at its equatorial plane.
The magnetic reconnection has the point-like reconnection region and
the slow shock waves, as in the Petschek reconnection model.
We also observed formation of magnetic islands by the
magnetic reconnection process.

In Section 2, we show basic equations of our resistive GRMHD code
and present assumptions of our simulations. Our
numerical simulation results are given in 
Section 3. We discuss and summarize our work in Section 4.

%% 4prl %% 
% \newpage

\section{Method of resistive GRMHD simulations
%\section{Basic equations
\label{sec2}}

\subsection{Covariant form of resistive GRMHD equations}

To investigate the basic process of magnetic reconnection around a
black hole, we performed
numerical simulations with the resistive GRMHD equations of 
plasmas and 
electromagnetic fields around a Schwarzschild black hole. 
In the space-time,
$x^\mu = (t, x^1, x^2, x^3) = (t, x^i) = (t, r, \theta, \phi)$
around a Schwarzschild black hole with mass $M$, the line element $ds$  
is given by 
\begin{equation}
ds^2 = g_{\mu \nu} dx^\mu dx^\nu = -\alpha^2dt^2 + \sum_{i}h^2_i(dx^i)^2
= -\alpha^2dt^2 + \frac{1}{\alpha^2}dr^2 + r^2d\theta^2 + 
r^2 {\rm sin}^2 \theta d\phi^2,
\end{equation}
where $\alpha = \sqrt{1 - r_{\rm S}/r}$ is the lapse function
and $r_{\rm S} = 2M$ is the Schwarzschild radius.
Here,
Greek subscripts such as $\mu$ and $\nu$ run from 0 to 3,
whereas Roman subscripts such as $i$ and $j$ run from 1 to 3.
We use the natural unit system where the speed of light $c$, the 
gravitational constant $G$, the magnetic permeability and electric
permittivity in a vacuum $\mu_0, \epsilon_0$, are unity. 

The covariant form of standard resistive GRMHD equations consists of the
conservation law of particle number, the conservation law of energy and
momentum, the Maxwell equations, and the 
relativistic Ohm's law with resistivity.
Here, we ignore radiation-cooling effects, plasma viscosity, and
self-gravity.  
The equations are written as
\begin{eqnarray}
& \displaystyle \nabla _\mu ( \rho U^\mu ) = 0  , \\ 
& \displaystyle \nabla _\mu T^{\mu \nu} = 0  , 
\label{eqenm0} \\
& \displaystyle \nabla _\mu F^{\mu \nu} = \frac{1}{\sqrt{-g}} \frac{\partial}{\partial x^\mu} 
\left ( \sqrt{-g} F^{\mu \nu} \right )= - J^\nu  ,
\label{eqam} \\
& \displaystyle  {\nabla_\mu} {^\ast}F^{\mu \nu} = \frac{1}{\sqrt{-g}} 
\partial _\mu ( \sqrt{-g}  {^\ast}F^{\mu \nu} ) = 0  , 
\label{eqfa_dual} \\
& \displaystyle F_{\mu \nu} U^\nu = \eta [J_\mu + ( U_\nu J^\nu ) U_\mu]  ,  
\end{eqnarray}
where $\nabla_\mu$ is the covariant derivative,
$\rho$ is the proper mass density,
$U^\mu$ is the four-velocity, $T^{\mu \nu}
= h U^\mu U^\nu + p g^{\mu \nu} + F^{\mu \sigma} {F^\nu}_\sigma -
(F^{\rho \sigma} F_{\rho \sigma}) g^{\mu \nu} / 4$
is the energy-momentum tensor of the plasma and the electromagnetic field
($h$ is the proper enthalpy density and $p$ is the pressure of the plasma),
$F_{\mu \nu}$ is the electromagnetic field tensor,
${^\ast}F_{\mu \nu} = \epsilon^{\mu \nu \rho \sigma} F_{\rho \sigma}/2$ is the 
dual tensor of $F_{\mu \nu}$,
%(\epsilon^{\mu \nu \rho \sigma}$ is the Levi-Civita antisymmetric tensor,
%which is a tensor density of weight $-1$),
$J^\mu =(\rho_{\rm e}, J^1, J^2, J^3)$
is the four-current density ($\rho_{\rm e}$ is the electric charge density),
and $\eta$ is the resistivity of the plasma.
$\epsilon^{\mu \nu \rho \sigma}$ is the Levi-Civita tensor, which is
defined as $\epsilon^{\mu \nu \rho \sigma} = 
\eta^{\mu \nu \rho \sigma}/\sqrt{-g}$,
$g = {\rm det} (g_{\mu \nu})$ 
is the determinant of $(g_{\mu \nu})$, and
$\eta^{\mu \nu \rho \sigma}$
is the totally asymmetric symbol defined as $\eta^{\mu \nu \rho \sigma} = 1$
if the order $[\mu \nu \rho \sigma]$ is an even permutation of $[0123]$,
$\eta^{\mu \nu \rho \sigma} = -1$
if the order $[\mu \nu \rho \sigma]$ is an odd permutation of $[0123]$,
and $\eta^{\mu \nu \rho \sigma} = 0$
unless $\mu, \nu, \rho, \sigma$ are all different.

\subsection{3+1 Formalism of Resistive GRMHD Equations}
In order to derive the 3+1 formalism of the resistive GRMHD equations, 
we introduce 
the local coordinate frame called the 
``zero angular momentum observer (ZAMO) frame,"
$\hat{x}^{\mu} = (\hat{t}, \hat{x}^i)$,
which is defined by  
\begin{equation}
ds^2 = \eta_{\mu \nu} d \hat{x}^{\mu} d \hat{x}^{\nu}
= - d \hat{t}^2 + \sum_{i=1}^3 (d \hat{x}^{i})^2  ,
\end{equation}
where $\eta_{\mu \nu}$ is the metric of Minkowski spacetime. 
Comparing Equations (1) and (7), we have the relation
$d\hat{t} = \alpha dt, d\hat{x}^i = h_i dx^i$.
We use the quantities observed by the ZAMO frame because they can be treated
intuitively, 
%and yield formulae more easily. 
because the relations
between the variables in the ZAMO frame are the same as those in the special
theory of relativity.
%and similar to the Newtonian relation.
Hereafter we denote
the variables observed in the ZAMO frame with the hat.

Using the quantities of the electromagnetic field in the ZAMO frame,
the resistive GRMHD equations are written in the following 3+1 formalism, 
\begin{eqnarray}
& \displaystyle 
\frac{\partial D}{\partial t} = - \frac{1}{h_1 h_2 h_3} \sum _i
\frac{\partial}{\partial x^i} \left (
%\frac{\alpha h_1 h_2 h_3}{h_i} D (\hat{v}^i + \beta ^i)
\frac{\alpha h_1 h_2 h_3}{h_i} D \hat{v}^i
\right )   , \\
& \displaystyle 
\frac{\partial \hat{P}^i}{\partial t} = - \frac{1}{h_1 h_2 h_3} \sum _j
\frac{\partial}{\partial x^j} \left ( 
%\frac{\alpha h_1 h_2 h_3}{h_j} (\hat{T}^{ij} + \beta ^j \hat{P}^i)
\frac{\alpha h_1 h_2 h_3}{h_j} \hat{T}^{ij} 
\right ) 
%\nonumber \\
%& \displaystyle 
- ( \epsilon + D) \frac{1}{h_i}
\frac{\partial \alpha}{\partial x^i} + \alpha f_{\rm curv}^i  , \\
%- \sum _j \hat{P}^j \sigma _{ji}
%+\sum_j \alpha \beta^j (G_{ij} \hat{P}^i -G_{ji} \hat{P}^j)  , \\
%\label{cmmo}
& \displaystyle 
\frac{\partial \epsilon}{\partial t} =  - \frac{1}{h_1 h_2 h_3} \sum _i
\frac{\partial}{\partial x^i} \left [
\frac{\alpha h_1 h_2 h_3}{h_i} (\hat{P}^{i} -D \hat{v}^i)
%+ \beta ^i \epsilon)
\right ]
%\nonumber \\
%& \displaystyle 
- \sum _i \hat{P}^i \frac{1}{h_i} \frac{\partial \alpha}{\partial x^i}  , \\
%-  \sum _{i,j} \hat{T}^{ij} \sigma _{ji}
%- \sum_{i} \alpha \beta^i  f_{\rm curv}^i  ,\\
%\label{cmem}
& \displaystyle 
\hat{E}_i + \hat{\epsilon} _{ijk} \hat{v}^j \hat{B}_k =
\frac{\eta}{\gamma} \left [ \hat{J}^i - \gamma ^ {2} ( \rho_{\rm{e}} -  
\hat{v}_j \hat{J}^j )  
\hat{v}^i \right ]  ,\\
& \displaystyle 
\frac{\partial \hat{B}_i}{\partial t} = \frac{- h_i}{h_1 h_2 h_3} \sum _{j,k}
\epsilon ^{ijk} \frac{\partial}{\partial x^j}
( \alpha h_k \hat{E}_k)  , \\ 
%\label{cmfa}
& \displaystyle 
\sum _{i} \frac{1}{h_1 h_2 h_3} \frac{\partial}{\partial x^i}
\left ( \frac{h_1 h_2 h_3}{h_i} \hat{B}_i
\right ) = 0  ,\\
& \displaystyle 
\rho _{\rm e} = \sum _{i} \frac{1}{h_1 h_2 h_3}
\frac{\partial}{\partial x^i} \left (
\frac{h_1 h_2 h_3}{h_i} \hat{E}_i
\right )  ,\\
%\label{dive}
& \displaystyle 
\alpha \hat{J}^i 
+ \frac{\partial \hat{E}_i}{\partial t} =
\sum _{j,k} \frac{h_i}{h_1 h_2 h_3} \epsilon ^{ijk}
\frac{\partial}{\partial x^j} ( 
\alpha h_k  \hat{B}_k )  , 
%\label{cmam}
\end{eqnarray}
where $D = \gamma \rho$ is  
special relativistic mass density, 
$\gamma$ is the Lorentz factor,
$\hat{v}^i = \hat{u}^i/\gamma$
is the three-velocity,  
$\hat{P}^i = h \gamma^2 \hat{v}^i + \hat{\epsilon}^{ijk} \hat{E}_j \hat{B}_k$ 
is the special
relativistic total momentum density, 
$\epsilon = \hat{T}^{00} - D = h \gamma^2 -p -D + 
%$\epsilon = \hat{T}^{00} - D = h \gamma^2 -P -D + 
\hat{B}^2/2 + \hat{E}^2/2$ is the special relativistic
total energy density, 
$\hat{T}^{ij} = h \gamma^2 \hat{v}^i \hat{v}^j + (p + \hat{B}^2/2 +
%$\hat{T}^{ij} = h \gamma^2 \hat{v}^i \hat{v}^j + (P + \hat{B}^2/2 +
\hat{E}^2/2) \delta^{ij} - \hat{B}^i \hat{B}^j - \hat{E}^i \hat{E}^j$
is the total stress tensor, 
$\hat{E}_i = \hat{F}_{i0}$ is the electric field, 
$\hat{B}^i = {^\ast}\hat{F}^{0i}$ is the magnetic field, and 
\[
f_{\rm{curv}}^{i} \equiv - \sum_{j} \left(
\frac{1}{h_{i} h_{j}} \frac{\partial h_i}{\partial x^j}
\hat{T}^{ij} - 
\frac{1}{h_j^2} \frac{\partial h_j}{\partial x^i}
\hat{T}^{jj} \right)   
\]
is the term containing \rm the centrifugal force.
Here, we used $\hat{\epsilon}^{ijk} \equiv \eta^{0ijk}$.

We solve these 3+1 form of equations numerically. 
We extended the numerical method of the resistive
special 
relativistic MHD (resistive RMHD) developed by Watanabe \& Yokoyama
(2006) to the general relativistic version. 
Note that Watanabe \& Yokoyama (2006) developed the resistive
RMHD code for the first time,
and they carried out numerical simulations of two-dimensional 
magnetic reconnection.  
Resistive RMHD simulations of magnetic reconnection were also presented by 
Zenitani et al. (2010), who discovered the post-plasmoid vertical shocks and
the diamond-chain structure.
We employ the
HLL flux solver 
and the MUSCL interpolation 
for the numerical simulation (Koide \& Morino 2011; Morino 2011).
We assume the plasma and field are axisymmetric with respect to the 
axis of the black hole.  

\subsection{
Setup of numerical resistive GRMHD simulations}

As the initial condition of the magnetic field,  
we have the split monopole magnetic field
around the Schwarzschild black hole:
\begin{equation}
\displaystyle
\hat B^r = 
\frac{B_0}{r^2} {\rm tanh} \left ( \frac{\theta - \pi / 2}{\Delta \theta_{\rm 
cw}}
\right ),
\verb!   !    
\hat B^{\theta} = \hat B^{\phi} = 0  , 
\end{equation}
where 
$B_0$ is a constant and
$\Delta \theta_{\rm cw}$ gives the 
current sheet width (thickness of the current sheet) at the 
equatorial plane. 
To introduce this split monopole magnetic field, we
refer to the antiparallel magnetic field given by 
Harris (1962).
We set the plasma and the magnetic field around the current sheet at the
equatorial plane and they are vertically in equilibrium initially.
The initial conditions of the plasma are given as
\begin{eqnarray}
& \displaystyle  \rho = \frac{\rho_{0}}{\sqrt{2Mr^{3}}}  ,\\
& \displaystyle p =  \frac{B^{2}_{0}}{2r^{4}} \frac{1}{\rm {cosh}^{2}[(
\theta - \pi/2) / \Delta \theta_{\rm cw}]} + p_{\rm b}  ,
\verb!   !    
p_{\rm b} = \frac{\beta_{\rm p}B^{2}_{0}}{8 \pi} \left (\frac{\rho}{\rho_{0}}
\right )^{\Gamma}  ,\\
& \displaystyle  \hat{v}^{r} = - 0.8 \sqrt{\frac{2M}{r}}  ,
\verb!   !    
\hat{v}^{\phi} = 
\hat{v}^{\theta} = 0  ,
\end{eqnarray}
where  
$\beta_{\rm p} \equiv p/(\hat{B}^2/2)$ is
%$\beta_{\rm p} \equiv P/(\hat{B}^2/2)$ is
the plasma beta value. 
We set the resistivity of the plasma 
$\eta$ uniform in space and constant in time in this paper.

With respect to the radial coordinate $r$,  
%We use the Boyer-Lindquist coordinate ($ct, r, \theta, \phi$).
we actually employ the modified tortoise coordinate,
$x = {\rm ln}[(r - r_{\rm min})/a_0 + 1]$. 
Here, $r_{\rm min}$ is the radial coordinate of the inner boundary near the
horizon and $a_0$ is a constant.
With a uniform mesh in the $x$-coordinate,
%along with the azimuthal $\phi$ and polar coordinates $\theta$,
the radial mesh width of the $r$-coordinate is proportional to
$r - r_{\rm min} + a_0$. Since the eigenspeed of the MHD waves near the black 
hole is very small because of the lapse function $\alpha$, the CFL
numerical stability condition is the most severe near $r = 1.5 r_{\rm S}$, 
while it is not 
severe near the black hole where the
mesh width is the smallest. 
This indicates that these modified tortoise coordinates ($x, \theta, 
\phi$) 
are appropriate
for the calculation both near and far from the black hole
(Koide et al. 1999).
We set  
the calculation region as $r_{\rm min} \leq r \leq 
r_{\rm max} = r_{\rm min} + a_0 [ (1 + \Delta \theta )^I -1 ],
\Delta \theta / 2 \leq \theta \leq \pi - \Delta \theta / 2$, 
where $r_{\rm min} = 1.001 r_{\rm S}, a_0 = 0.4,  
\Delta \theta = \pi (1 - 1 / J ) / J $
is the mesh width of $\theta$, and $I, J$ are grid numbers  
for $r$ and $\theta$, respectively. 
We have the radial mesh
width $\Delta r = ( r - r_{\rm min} + a_0) \Delta \theta$.
Then, the minimum radial mesh width is given at $r = r_{\rm min}$ as 
$\Delta r_{\rm min} = a_0 \Delta \theta$.
For the numerical calculations, we choose
$I, J, \Delta \theta, r_{\rm max} (\Delta \theta$ is calculated from $J$
and $r_{\rm max}$ is calculated from $\Delta \theta$ and $I$), and 
the time interval $\Delta t$,
for different $\eta$ values as 
shown in Table 1.
The numerical stability 
conditions are given by 
(i) $\Delta t \leq ( \Delta r / \alpha)_{\rm min}$,  
\footnote{$1 \leq ( \Delta r / \alpha)_{\rm min} / \Delta t
\leq \Delta r / (\alpha \Delta t) =  
r \Delta \theta / (\alpha \Delta t)$. 
%< r \Delta \theta / (\alpha \Delta t)$.
}
and (ii)
$\Delta t < 2\eta$. 
\footnote{
From the Ampere's law, 
$\displaystyle
\frac{\partial} 
{\partial t} \vec{E} = - \alpha \vec{J}$,  
and from the Ohm's law, $\vec{E} = \eta \vec{J}$, we have $\displaystyle  
\frac {\partial}{\partial t} \vec{E} = - \frac{\alpha \vec{E}}{\eta},
\frac {1}{\Delta t}(\vec{E}^{n+1} - \vec{E}^n) = - \frac{\alpha 
\vec{E}^n}{\eta}$, and
$\vec{E}^{n+1} = (1 - \alpha \Delta t/\eta) \vec{E}^n$.
The numerical stability condition is $-1 < (1 - \alpha \Delta t/\eta)$, 
i.e., $2\eta > \Delta t > \alpha \Delta t$.}
The $\eta, \Delta \theta$, and $\Delta t$ 
combinations in Table 1 
satisfy these conditions 
%($\eta$ and $\Delta t$ sets actually fill $\Delta t / \eta < 1/2$
%in these combinations).
(these $\eta$ and $\Delta t$ satisfy $\Delta t / \eta < 1/2$).

%\begin{table}[t]
\begin{table}[h]
\caption{
Numerical Conditions of Resistive GRMHD Simulations}
\begin{center}
\begin{tabular}{|c|c|c|c|c|c|c|} \hline
$\eta / r_{\rm S}$ & $I$ & $J$ &
$r_{\rm min}$ & $r_{\rm max}$ & $\Delta \theta$ & $\Delta t / \tau_{\rm S}$ \\
\hline \hline
$5 \times 10^{-3}$ & 450 & 216 & 1.001$r_{\rm S}$ & 87.8 &
$1.20 \times 10^{-2}$ &
$5 \times 10^{-4}$  \\
$3 \times 10^{-3}$  &  &  &  &  &  &   \\
\cline{1-3} \cline{5-6}
$2 \times 10^{-3}$ & 600 & 288 &  & 263  & $1.09 \times 10^{-2}$  &   \\
$1 \times 10^{-3}$  &  &  &  &  &  &   \\
\cline{1-3} \cline{5-7}
$5 \times 10^{-4}$ & 1200 & 576 &  & 271  & $5.44 \times 10^{-3}$  &
$1 \times 10^{-4}$  \\
\cline{1-1} \cline{7-7}
$3 \times 10^{-4}$ &  &  &  &  &  & $5 \times 10^{-5}$ \\
$1 \times 10^{-4}$  &  &  &  &  &  &   \\
\cline{1-3} \cline{5-7}
$5 \times 10^{-5}$ & 1800 & 864 &  & 274  & $3.63 \times 10^{-3}$  &
$1 \times 10^{-5}$  \\
\cline{1-1} \cline{7-7}
$1 \times 10^{-5}$  &  &  &  &  &  & $5 \times 10^{-6}$  \\
\hline
\end{tabular}
%\caption{
%%Numerical conditions of resistive GRMHD simulations.
%The calculation region is $r_{\rm min} \leq r \leq
%r_{\rm max} = r_{\rm min} + a_0 [ (1 + \Delta \theta )^I -1 ],
%\Delta \theta / 2 < \theta < \pi - \Delta \theta / 2$,
%where $r_{\rm min} = 1.001 r_{\rm S}, a_0 = 0.4,
%\Delta \theta = \pi (1 - 1 / J ) / J $
%is the mesh width of $\theta$, and $I, J$ are grid numbers
%for $r$ and $\theta$.
%The radial mesh
%width is $\Delta r = ( r - r_{\rm min} + a_0) \Delta \theta$.
%Then, the minimum radial mesh width is given at $r = r_{\rm min}$ as
%$\Delta r_{\rm min} = a_0 \Delta \theta$.
%We choose
%$I, J, \Delta \theta, r_{\rm max} (\Delta \theta$ is calculated from $J$
%and $r_{\rm max}$ is calculated from $\Delta \theta$ and $I$), and
%the time interval $\Delta t$
%for different $\eta$ values as
%shown in this table in the simulations.
%The combinations of
%$\eta, \Delta \theta$, and $\Delta t$ in this table satisfy
%the numerical stability conditions mentioned in the text.
%%\label{}}
\end{center}
\end{table}
%\hspace{1mm}
%\hspace{5mm}
\footnotesize \bf Note. \rm The calculation region is $r_{\rm min} \leq r \leq
r_{\rm max} = r_{\rm min} + a_0 [ (1 + \Delta \theta )^I -1 ],
\Delta \theta / 2 < \theta < \pi - \Delta \theta / 2$,
where $r_{\rm min} = 1.001 r_{\rm S}, a_0 = 0.4,
\Delta \theta = \pi (1 - 1 / J ) / J $
is the mesh width of $\theta$, and $I, J$ are grid numbers
for $r$ and $\theta$.
The radial mesh
width is $\Delta r = ( r - r_{\rm min} + a_0) \Delta \theta$.
Then, the minimum radial mesh width is given at $r = r_{\rm min}$ as
$\Delta r_{\rm min} = a_0 \Delta \theta$.
We choose
$I, J, \Delta \theta, r_{\rm max} (\Delta \theta$ is calculated from $J$
and $r_{\rm max}$ is calculated from $\Delta \theta$ and $I$), and
the time interval $\Delta t$
for different $\eta$ values as
shown in this table in the simulations.
The combinations of
$\eta, \Delta \theta$, and $\Delta t$ in this table satisfy
the numerical stability conditions mentioned in the text.
\normalsize

\newpage
\section{Numerical simulation results}

We present the simulation results with the initial conditions of $
\rho_{0} = 1, B_{0} = 10, \Delta \theta_{\rm CW} = 0.1$, and
$\beta_{\rm p} = 0.025$.
%We use the Schwarzschild transit time   
%$\tau_{\rm S} = r_{\rm S}$ as units of time.

Resistivity $\eta$ values are set from $1 \times 10^{-5} r_{\rm S}$ to
$0.005 r_{\rm S}$. Here we convert these $\eta$ values to the magnetic
Reynolds numbers. The magnetic Reynolds number
$S$ is defined as $S = Lv / \eta$ in MHD, 
where $L, v$ are typical length-scale and
velocity of plasma. In the Sweet--Parker and Petschek
mechanisms, we identify $L$
(the length of the reconnection sheet) with the global external length-scale
$L_{\rm e}$ and $S$ therefore with the global magnetic Reynolds number 
$S_{\rm e} = L_{\rm e} v_{\rm A} / \eta$, where $v_{\rm A}$ is the Alfven 
velocity of the plasma
(Parker 1957; Sweet 1958; Petschek 1964;
Priest \& Forbes 2000).
The relativistic Alfven velocity is calculated as 
$v_{\rm A} = 
\sqrt{\bar{B}^2/(\bar{h} + \bar{B}^2)}$, where
$\bar{B}$ and $\bar{h}$ are typical values of the magnetic flux density 
in the ZAMO frame 
and the proper
enthalpy density, respectively, outside of the current sheet.
Substituting 
$\bar{B}
\sim 10$ and $\bar{h} \sim \rho_{0} = 1$ from the initial conditions, we
obtain $v_{\rm A} \sim 1$. 
%obtain $v_{\rm A} \approx 1$. 
If we regard $L_{\rm e} = r_{\rm S}$, then $S_{\rm e} = r_{\rm S}/\eta$ with
our initial conditions. Thus $\eta = 1 \times 10^{-5} r_{\rm S}$ 
corresponds to $S_{\rm e}
= 10^5$ and $\eta = 
0.005 r_{\rm S}$ corresponds to $S_{\rm e} = 200$ in our simulations.

\begin{figure} % [H]
\begin{center}
\includegraphics[width=12cm]{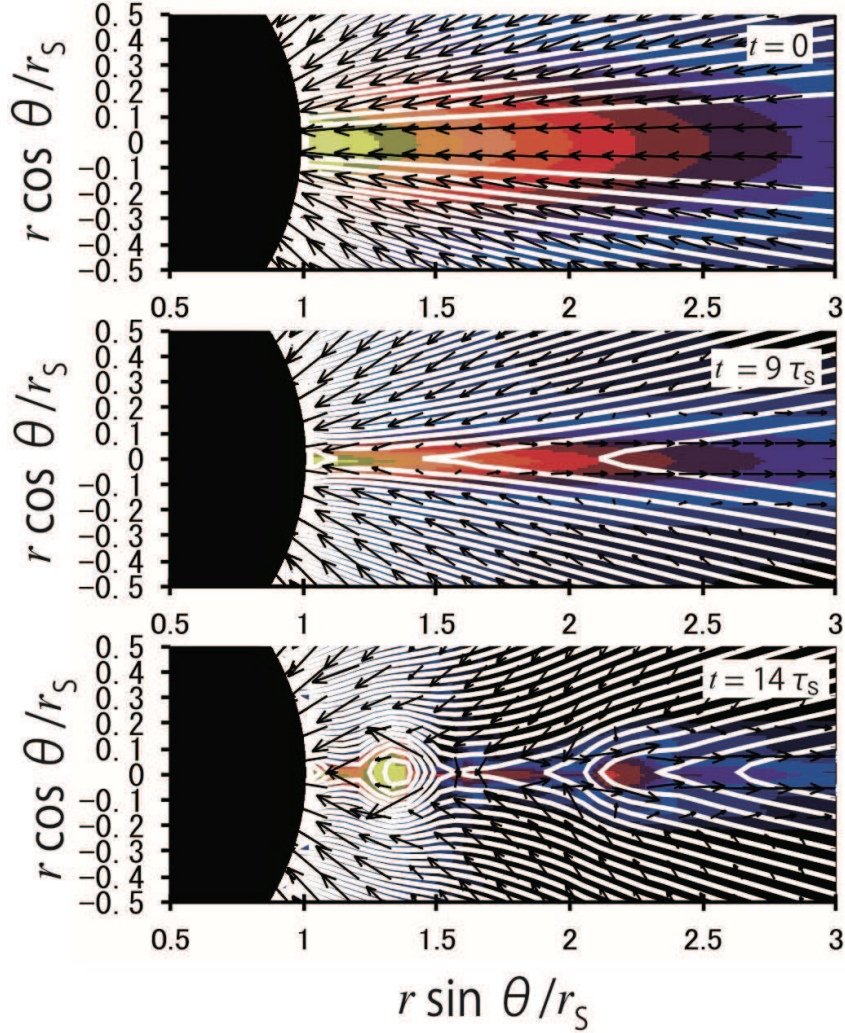}
% Here is how to import EPS art
\caption{
Time evolution of pressure (color), magnetic field (white lines), and velocity
(black arrows) by the resistive GRMHD simulations with the resistivity
$\eta = 0.001 r_{\rm S}$ or the global magnetic Reynolds number
$S_{\rm e} = 10^3$.
Black regions on the left in the panels show the
horizon of the black hole.
Top panel: plasma and magnetic field at the initial state.
The initial magnetic field is the
split monopole type; above the equatorial plane
of the black hole, the magnetic field lines are directed toward the black
hole, and below the equatorial plane the field lines are directed away from
the black hole.
Middle panel: at $t = 9 \tau_{\rm S}$, single magnetic reconnection occurs
around the current sheet near the horizon at $r \sim 1.2 r_{\rm S},
\theta = \pi / 2$.
The reconnection region is
point-like and narrow, and the slow shock waves are seen, as the Petschek
reconnection model.
Bottom panel: at $t = 14 \tau_{\rm S}$, multiple reconnections happen and
the plasmoid is formed around
$r \sim 1.4 r_{\rm S}, \theta = \pi / 2$.
\label{fig2}}
\end{center}
\end{figure}

Figure 2 shows the time evolution of pressure (color), magnetic field
(white lines), and velocity (arrows) in the case of $\eta = 0.001 r_{\rm S}$. 
$\eta = 0.001 r_{\rm S}$ corresponds to  
the global magnetic
Reynolds number $S_{\rm e}  = 10^3$.
Black regions at the left in Figure 2 
show the horizon of the black hole. 
Initially the plasma and the magnetic field around the current sheet at the
equatorial plane are vertically in equilibrium (Figure 2, top panel). 
The initial magnetic field is split monopole type;
above the equatorial plane
of the black hole, the magnetic field lines are directed toward the black
hole, and below the equatorial plane the field lines are directed away from
the black hole.
At $t = 9
\tau_{\rm S}$, the magnetic reconnection occurs around the current sheet
near the horizon
at $r \sim 1.2 r_{\rm S}, \theta = \pi /2$ 
(Figure 2, middle panel). 
The reconnection region seems point-like and narrow, and the
slow shocks are found along the current sheet outside of the reconnection point,
which is similar to the Petschek reconnection model. At $t = 14 \tau_{\rm S}$,
multiple magnetic reconnections are caused and the plasmoid is formed around
$r \sim 1.4 r_{\rm S}, \theta = \pi / 2$ (Figure 2, bottom panel).  
Because the resistivity is set to be
uniform, this relatively fast
magnetic reconnection, as in the Petschek model,
was not expected until we saw this simulation result.
There are arguments, as we mentioned in section 1,  
that even though the plasma resistivity is spatially uniform, 
Petschek-type fast stationary magnetic reconnection is achieved
through the use of a nonuniform viscosity profile
%solution could possibly exist
%when other physical profiles,
%e.g., viscosity distribution, are nonuniform,
%or obstacles in the flow along the diffusion region are used
(Baty et al. 2009). 
In the present case, however, viscosity is assumed to be zero, while 
the thickness of the current sheet and the lapse
of time are nonuniform (see also section 4). 
The time lapse described by $\alpha$ 
is the general relativistic effect.  
Thus the general relativistic effect plays an important role for the 
magnetic reconnection near the black hole.

We observed the magnetic reconnection rate at the reconnection point for a
given $\eta$ at a given time of the resistive GRMHD simulations. 
We define the diffusive slip-through rate of
magnetic field lines across plasma at any point as 
\begin{equation}
\displaystyle  R_{\rm ms} = \frac{\alpha E'_{\phi}}{v_{\rm A}\bar{B}}
=  \frac{\alpha \eta \hat{J}_{\phi}}{v_{\rm A}\bar{B}},
\end{equation}
where $E'_{\phi} \equiv \eta \hat{J}_{\phi}$
is the electric field measured by the plasma rest frame.
%\bf In equation (20), $\alpha$ is the general relativistic term. 
$\bar{B}$ is the magnetic field strength  
just above
the reconnection
point, and outside of the current sheet. 
Without $\alpha$, this equation is the 
definition of the standard (nonrelativistic) magnetic reconnection rate. 
To find out the location of the reconnection point,
we checked the profile of
$-\hat{B}_{\theta}$
along the equatorial plane (Figure 3, top). Because $\hat{B}_{\theta}$
vanishes
at the reconnection point, we identify the position of reconnection
point $r = r_{\rm X}$
by the position of $\hat{B}_{\theta} = 0$ at the equatorial plane. 
The magnetic 
reconnection rate, $R_{\rm mr}$, is given by $R_{\rm ms}$ at the reconnection
point (Figure 3, middle). The radial component of the velocity  
$\hat{v}_{r}$ is negative at the reconnection
point (Figure 3, bottom), which means plasma is falling into the black hole at
the reconnection point.

\begin{figure} % [H]
\begin{center}
\includegraphics[width=7cm]{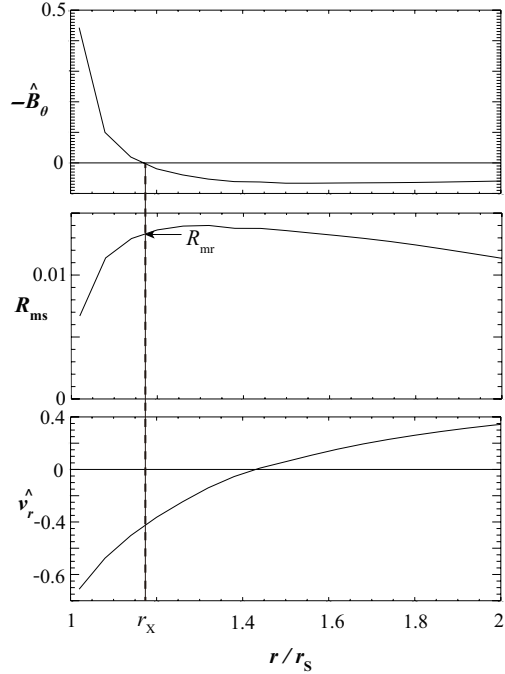}
\caption{
Azimuthal component of the magnetic field $(-\hat{B}_{\theta}),$ diffusive
slip-through rate of magnetic field lines across plasma $(R_{\rm ms})$, and
radial component of velocity $(\hat{v}_{r})$ as functions of $r / r_{\rm S}$
along the
equatorial plane, at $t = 9 \tau_{\rm S}$ in the case of
$\eta = 0.001 r_{\rm S}$. $r_{\rm X}$ is the position of the reconnection point.
This figure explains how to determine $r_{\rm X}$ (the position of
$\hat{B}_{\theta} = 0$) and the reconnection rate $R_{\rm mr} (R_{\rm ms}$
value at $r_{\rm X})$.
\label{fig3}}
\end{center}
\end{figure}

\begin{figure} % [H]
\begin{center}
\includegraphics[width=11cm]{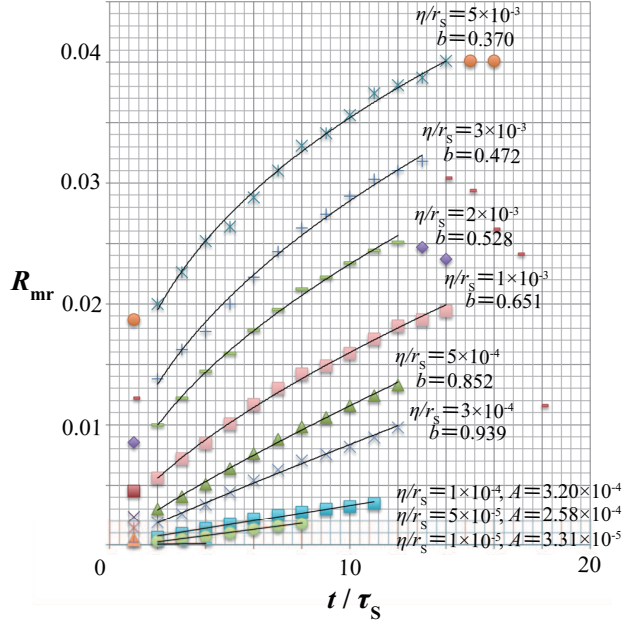}
\caption{
Time evolutions of magnetic reconnection rate $R_{\rm mr}$ for various
resistivity $\eta$ values in the range from
$1 \times 10^{-5} r_{\rm S}$ to $0.005r_{\rm S}$.
From $\eta = 3 \times 10^{-4}$ to $0.005 r_{\rm S}$, we fit the data with
$R_{\rm mr} = a (t/\tau_{\rm S})^b$, where $a$ and $b$ are constants,
from $t = 2 \tau_{\rm S}$ to
$\sim 10 \tau_{\rm S}$.
$\eta$ and the best-fit
$b$ values are shown in the figure, together with the best-fit lines.
For $\eta = 1 \times 10^{-5} r_{\rm S},
5 \times 10^{-5} r_{\rm S}$,
and $1 \times 10^{-4} r_{\rm S},
R_{\rm mr}$ can be represented by
$R_{\rm mr} = A(t/\tau_{\rm S}) + C$, where $A$ and $C$ are constants.
$\eta$, the best-fit $A$ values,
and the best-fit lines
are in the figure.
Orange circle and blue St. Andrew's cross: $\eta = 5 \times 10^{-3} r_{\rm S}$,
red rectangle and blue Greek cross: $\eta = 3 \times 10^{-3} r_{\rm S}$,
purple diamond shape and yellow-green rectangular:
$\eta = 2 \times 10^{-3} r_{\rm S}$,
red square and pink square: $\eta = 1 \times 10^{-3} r_{\rm S}$,
purple St. Andrew's cross and yellow-green triangle:
$\eta = 5 \times 10^{-4} r_{\rm S}$,
orange St. Andrew's cross and blue St Andrew's cross:
 $\eta = 3 \times 10^{-4} r_{\rm S}$,
orange triangle and blue square: $\eta = 1 \times 10^{-4} r_{\rm S}$,
purple Greek cross and yellow-green circle: $\eta = 5 \times 10^{-5} r_{\rm S}$,
orange rectangle and blue rectangle: $\eta = 1 \times 10^{-5} r_{\rm S}$.
\label{fig4}}
\end{center}
\end{figure}

\begin{figure} % [H]
\begin{center}
\includegraphics[width=10cm]{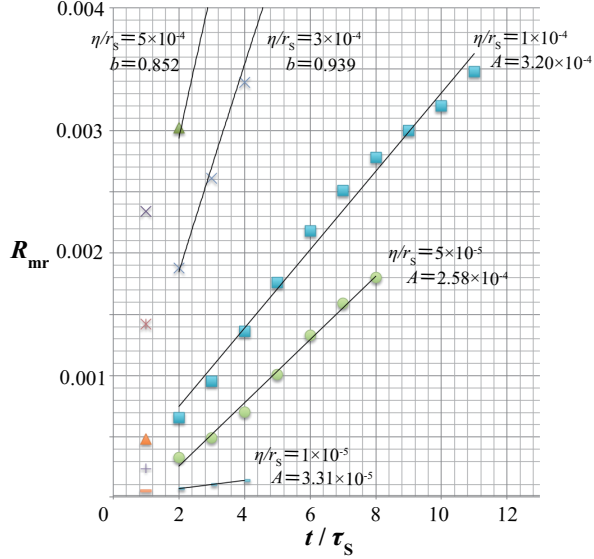}
\caption{
Same as Figure 4 but with scales of $0 \leq t/\tau_{\rm S} \leq 13$ and
$0 \leq R_{\rm mr} \leq 0.004$ to highlight the three low $\eta$ cases:
$\eta = 1 \times 10^{-5} r_{\rm S},
5 \times 10^{-5} r_{\rm S},$ and $1 \times 10^{-4} r_{\rm S}$.
\label{fig5}}
\end{center}
\end{figure}

\begin{figure} % [H]
\begin{center}
\includegraphics[width=9cm]{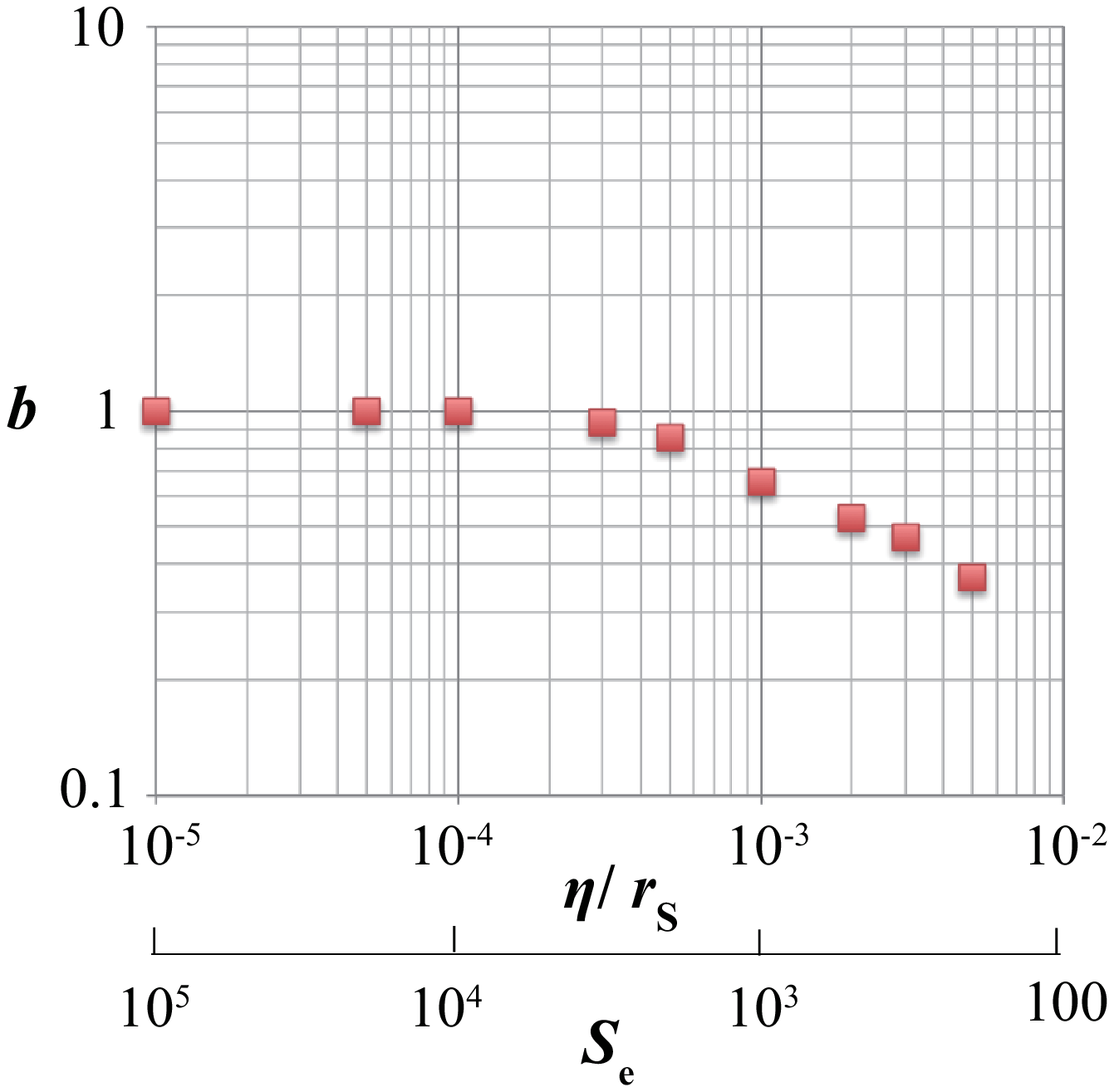}
\caption{
Dependence of the power index $b$ of $R_{\rm mr}$
time-development
(i.e., $R_{\rm mr} \propto t^b$)
on the resistivity $\eta$
in the range from $1 \times 10^{-5} r_{\rm S}$
to 0.005$r_{\rm S}$.
$b$ approaches to unity as $\eta$ becomes smaller, or $S_{\rm e}$
gets larger.
When $\eta \leq 1 \times 10^{-4} r_{\rm S}$ or
$S_{\rm e} \geq 10^4,
R_{\rm mr}$ increases linearly as
$t$ (see Figures 4 and 5) so
we set $b = 1$ for $\eta \leq 1 \times 10^{-4} r_{\rm S}$.
From the initial conditions,
$\eta = 10^{-5} r_{\rm S},
10^{-4} r_{\rm S}, 10^{-3} r_{\rm S}$, and
$10^{-2} r_{\rm S}$
correspond to the global
magnetic Reynolds number $S_{\rm e} = 10^5, 10^4, 10^3,$ and 100,
respectively. These $S_{\rm e}$ values are also shown
below the horizontal axis.
\label{fig6}}
\end{center}
\end{figure}

In Figures 4 and 5 we present 
time developments of magnetic reconnection rate $R_{\rm mr}$
for various resistivity $\eta$ values in the range from
$1 \times 10^{-5} r_{\rm S}$ to 
0.005$r_{\rm S}$.
%Note that 
%$\eta = 10^{-5} r_{\rm S},
%10^{-4} r_{\rm S},$ and 
%$0.001 r_{\rm S}$ 
%correspond to the magnetic Reynolds number  
%$S = 10^4, 10^3,$ and 100, respectively, from the initial
%conditions. 
As shown in Figure 4, $R_{\rm mr}$
decreases as 
$\eta$ becomes smaller.
Regarding the time-development, 
within the $\eta$ range of $3 \times 10^{-4} r_{\rm S}$
to $0.005 r_{\rm S}$,  
$R_{\rm mr}$
is a function of powers of time,
$R_{\rm mr} \propto t^b$ ($b$ is a constant)
from $t = 2 \tau_{\rm S}$ to $\sim 10 \tau_{\rm S}$,
and it tends to settle down to a constant value
afterwards for large resistivity.
For $\eta = 1 \times 10^{-5} r_{\rm S}, 5 \times 10^{-5} r_{\rm S}$,
and $1 \times 10^{-4}
r_{\rm S}$, we fit $R_{\rm mr}$ by a linear function of $t$, 
$R_{\rm mr} = A(t/\tau_{\rm S}) + C$,
where $A$ and $C$ are constants. The results of these three low $\eta$ cases
are highlighted in Figure 5. $R_{\rm mr}$ values of these low $\eta$
cases are well fit by the 
linear functions of $t$. 
Regrettably, low-$\eta$ simulations ($\eta = 5 \times 10^{-5} r_{\rm S}$ and 
especially
$1 \times 10^{-5} r_{\rm S}$) run only for short times as shown in Figures 4
and 5,
because of numerical difficulty in the cases of low-$\eta$. 
%the resistive GRMHD equations with low-$\eta$ are numerically unstable. 
It takes a lot of time for low-$\eta$ runs to complete the job
because we should select very large mesh numbers ($I, J$), 
a very small time interval $\Delta t$, 
and an appropriate combination of them 
for low-$\eta$ calculations. 
The
runs often stop due to poor convergence for low-$\eta$ cases.  
We will improve this problem by establishing a more stable scheme of the
resistive GRMHD code
in the near future.
 
The relationship between the resistivity $\eta$ and the power index $b$
is offered by Figure 6. 
Power index $b$ approaches to unity as $\eta$ becomes smaller or $S_{\rm e}$ 
gets larger, and when  
$\eta \leq 
10^{-4} r_{\rm S}$ or $S_{\rm e} \geq 10^4$, 
$R_{\rm mr}$ is a linear function of $t$, 
which means $b = 1$ (see Figures 4 and 5). 
As we described in section 1 and will mention     
in section 4, among models of magnetic reconnection, only 
the Rutherford regime of the  
tearing mode instability explains the results of  
$R_{\rm mr} \propto t$ for small resistivity (Rutherford 1973).

\begin{figure} % [H]
\begin{center}
\includegraphics[width=8cm]{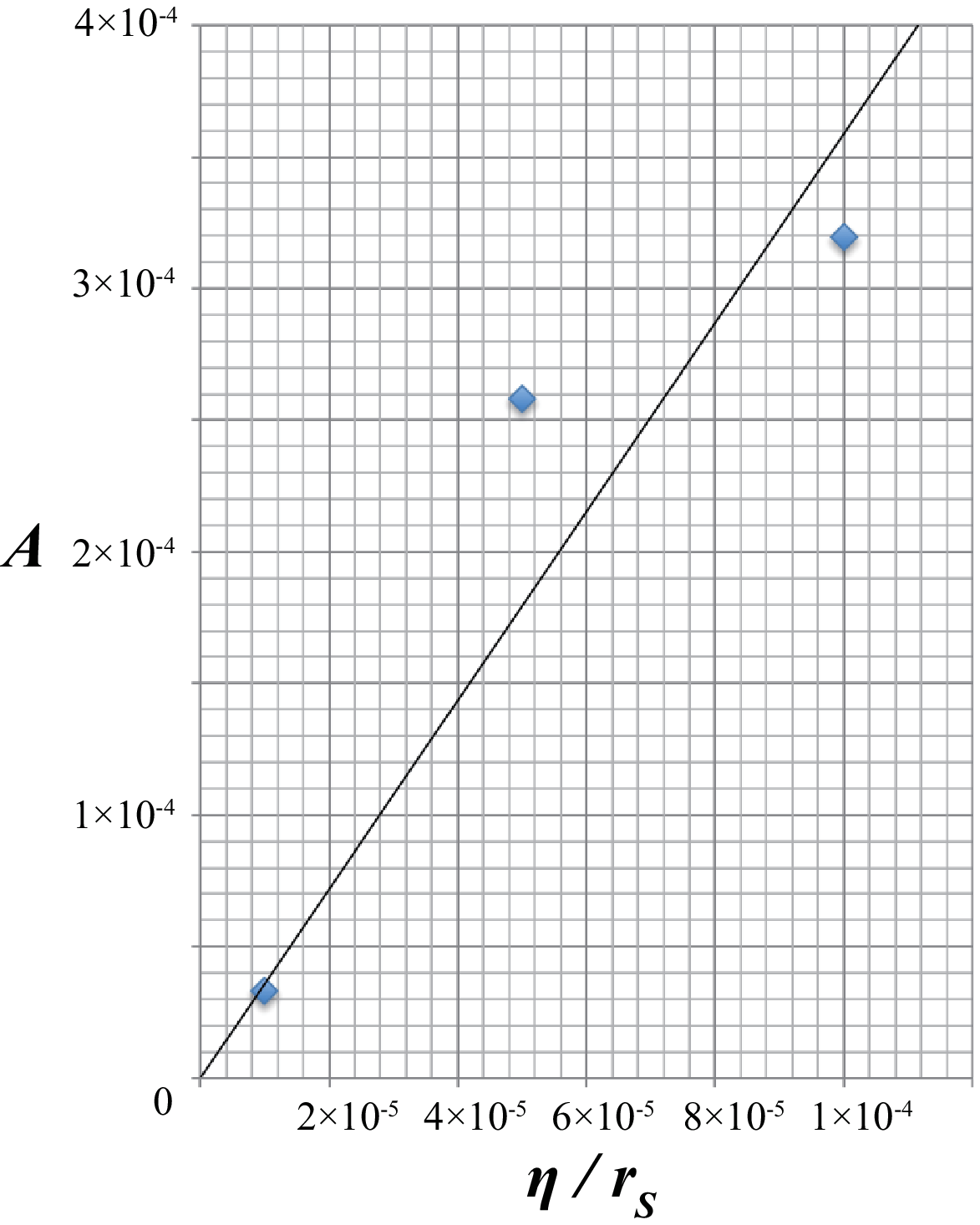}
\caption{
Dependence of time change rate of the magnetic reconnection rate
$R_{\rm mr}$ on the resistivity $\eta$ for
$\eta = 1 \times 10^{-5} r_{\rm S},
5 \times 10^{-5} r_{\rm S},$ and $1 \times 10^{-4} r_{\rm S}$.
Here
$R_{\rm mr}$ is fit by
$R_{\rm mr} = A(t/\tau_{\rm S}) + C$, where $A$ and $C$ are constants
(see also Figures 4 and 5).
This plot shows
$R_{\rm mr} \propto \eta$
(although the results are somewhat skewed),
which gives the evidence that the magnetic reconnection is
in the Rutherford regime of the tearing mode instability.
\label{fig7}}
\end{center}
\end{figure}

The theory of the Rutherford regime of the tearing mode instability also 
predicts $R_{\rm mr} \propto \eta$ (Rutherford 1973; Park et al. 1984;
Priest \& Forbes 2000),
which leads to $R_{\rm mr} \propto t \eta$. 
We plot the relationship between $\eta$ and $A$ 
for $\eta = 1 \times 10^{-4} r_{\rm S}, 
5 \times 10^{-5} r_{\rm S}$,
and $1 \times 10^{-5} r_{\rm S}$
in Figure 7. 
Figure 7 indicates   
$R_{\rm mr} \propto
\eta$, which gives the evidence that 
the magnetic reconnection we see with these $\eta$  
is the Rutherford regime of the  
tearing mode instability.

Whether the reconnection point $r_{\rm X}$
shifts with time or not is an interesting 
topic. The time evolutions of 
$r_{\rm X}$
for four 
different resistivity values can be seen in Figure 8.
For all the resistivity values
searched,
$r_{\rm X}$
stays almost at the same position at $r \sim 1.2
r_{\rm S}$ during $t \sim$ 5--12$\tau_{\rm S}$. 
In the earlier phase   
$(t < 5 \tau_{\rm S})$ and in 
the later phase $(t > 12
\tau_{\rm S})$, 
%the reconnection point
$r_{\rm X}$
approaches toward the black hole horizon.
The reason of this $r_{\rm X}$ motion can be understood as follows. 
We assumed the initial condition of the plasma velocity to be $\hat{v}^r < 0$
(Equation (19)), so 
in the earlier phase, the plasma initially falls into the black hole,
which moves $r_{\rm X}$ 
toward the black hole. During 
$t \sim$ 5--12$\tau_{\rm S}$, the single magnetic reconnection happens, and
the plasma ejection from the reconnection point stops the infall of the 
plasma, thus 
$r_{\rm X}$ stays at the same position at 
$r \sim 1.2 r_{\rm S}$. In the later phase, multiple magnetic reconnections
occur, making a magnetic island,  
and the magnetic reconnection nearest the black hole is isolated from 
the outer magnetic field lines thus $r_{\rm X}$ 
moves again toward the black hole.

\begin{figure} % [H]
\begin{center}
\includegraphics[width=9cm]{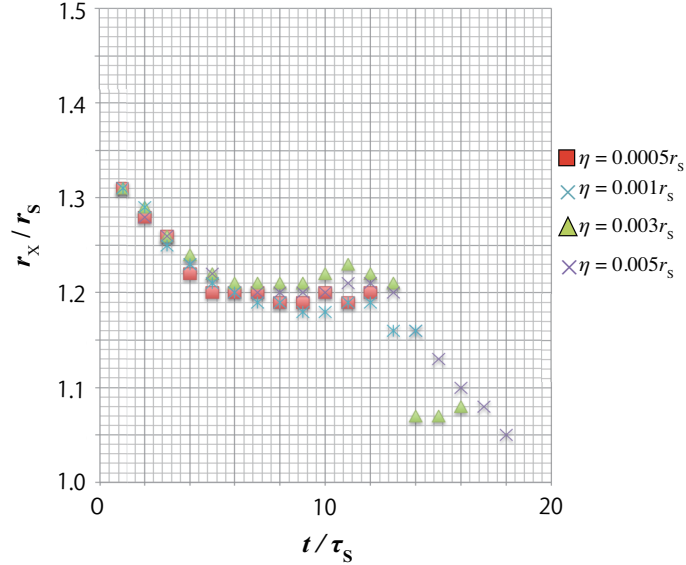}
\caption{
Time evolutions of reconnection point $r_{\rm X}$ for four
different resistivity
$\eta$ values.
Red square: $\eta = 5 \times 10^{-4} r_{\rm S}$,
blue St. Andrew's cross: $\eta = 1 \times 10^{-3} r_{\rm S}$,
yellow-green triangle: $\eta = 3 \times 10^{-3} r_{\rm S}$,
purple St. Andrew's cross: $\eta = 5 \times 10^{-3} r_{\rm S}$.
\label{fig8}}
\end{center}
\end{figure}

\newpage
\section{Discussion}  

To investigate basic physical process of magnetic reconnection
around a black hole, 
we have developed a resistive GRMHD code, and
performed numerical simulations of 
plasmas and electromagnetic field around a 
Schwarzschild black hole.   
We assumed  
split-monopole magnetic field around the black hole as the initial condition,
and electric resistivity $\eta$ to
be uniform in space and constant in time.   
We obtained the following results.
\begin{itemize}
\item We found that relatively
fast magnetic reconnection happens near the black hole at its
equatorial plane. 
This is the first resistive GRMHD simulation result that shows
clear magnetic reconnection feature in plasma around a black hole,
to our knowledge.     
The structure of the
reconnection is like the Petschek reconnection model, which has the
point-like reconnection region and the slow shock waves, 
while $\eta$ is assumed to be
uniform. We also observed formation of magnetic islands 
by the magnetic reconnection process. 
\item The magnetic reconnection rate $R_{\rm mr}$ decreases
as $\eta$ becomes smaller.
For $\eta > 1 \times 10^{-4} r_{\rm S}$, 
$R_{\rm mr}$ is a function of powers of time $t, R_{\rm mr} \propto t^b$
($b$ is the power index),
from $t = 2 \tau_{\rm S}$ to 
$\sim 10 \tau_{\rm S}$, 
and $R_{\rm mr}$ tends to settle down to a constant value afterwards for 
large $\eta$.
For $\eta \leq 1 \times 10^{-4} r_{\rm S}$ or
the global magnetic Reynolds number $S_{\rm e} \geq 10^4, 
R_{\rm mr}$ increases linearly as time.
For this range of $\eta$ or $S_{\rm e}$,
$R_{\rm mr}$ is also proportional to $\eta$, which
means $R_{\rm mr} \propto \eta t$. 
These results are in good agreement with the magnetic
reconnection in 
the Rutherford regime of the tearing mode instability. 
\end{itemize}

We discuss here the relationship between
present magnetic reconnection models
and our simulation results. 
Note that in astrophysical situations,
where plasma is very thin, the magnetic Reynolds
number, which is proportional to the inverse of resistivity, is supposed to
be sufficiently large ($S \gg 1)$.
Recent nonrelativistic MHD theories   
of magnetic reconnection 
with sufficiently large $S$ and initially uniform current sheet  
suggest the
form of time development of magnetic reconnection rate $R_{\rm mr}$
as shown in Figure 1 (e.g., Murphy et al. 2008). 
The models of magnetic reconnection 
can be classified into three periods 
in the time
dependence of $R_{\rm mr}$: 
the linear growth stage, 
the nonlinear growth stage, and the stationary stage.  
The former two stages come from the tearing mode instability.  
In the linear growth stage, 
the tearing mode instability starts to appear, 
and the time dependence of $R_{\rm mr}$ is
$R_{\rm mr} \propto e^{\gamma t}$, where $\gamma \propto
1 / \sqrt{S}$
(Furth et al. 1963; Biskamp 1993).
This exponential growth in time is considered as the beginning phase of 
magnetic reconnection.
After the linear growth stage, magnetic reconnection enters to the 
nonlinear growth stage. The phenomena in this stage is 
rather complex. The Rutherford
regime is known  
as the nonlinear growth stage of the 
tearing mode instability, and according to this theory, 
$R_{\rm mr} \propto t$
(Rutherford 1973). 
This can be regarded as the transition stage of 
magnetic reconnection.
Our results are in good agreement with this model during $t \sim 2-10 
\tau_{\rm S}$. 
If the magnetic reconnection is in the Rutherford regime of the tearing
mode instability, $R_{\rm mr}$ is also
proportional to $\eta$ (Rutherford 1973; Park et al. 1984;
Priest \& Forbes 2000).
Our results also show $R_{\rm mr} \propto \eta$ for $\eta \leq
1 \times 10^{-4}
r_{\rm S}$, which
agrees with this theoretical prediction. This confirms that the
magnetic reconnection we see in the simulation results
for $\eta \leq 1 \times 10^{-4} r_{\rm S}$ is
the phenomena in this regime. 
Finally, the steady-state reconnection is achieved
as the Sweet--Parker model or the Petschek model. 
The stationary models
predict $R_{\rm mr}$ to be constant
in time (time-independent), and this must be the final stage of magnetic
reconnection. 
The Sweet--Parker model leads $R_{\rm mr} = 1 / \sqrt{S_{\rm e}}$, whereas
the Petschek model tells 
$R_{\rm mr}$(max)$ = \pi / (8 {\rm ln} S_{\rm e})$.
Here 
$S_{\rm e}$ is the global magnetic Reynolds number and
$R_{\rm mr}$(max) is the maximum magnetic reconnection rate estimated
by Petschek, as explained
in Section 3  
(Parker 1957; Sweet 1958; Petschek 1964;  
Priest \& Forbes 2000; Kulsrud 2005; Shibata \& Magara 2011).

We mentioned above that magnetic reconnection starts from the linear growth 
stage (Figure 1), while 
our resistive
GRMHD simulations showed that magnetic reconnection starts from the Rutherford
regime, which is the second stage of the time-evolution of magnetic
reconnection. This
is explained by the break-down of the 
uniformity of the current sheet around the equatorial plane of the black hole. 
The current sheet is thinner as $r$ is 
smaller in the 
initial condition with 
the split-monopole magnetic field.
Then the current density becomes larger as $r$ gets smaller. 
However, at the horizon, $\alpha$ vanishes, then 
the diffusive slip-through rate of magnetic field lines across plasma 
$R_{\rm ms} = \alpha \eta \hat{J}^\phi / (v_{\rm A} \bar{B})$ has its maximum 
value outside of the horizon. The maximum $R_{\rm ms}$ point is 
expected to become the reconnection point $r_{\rm X}$. 
We examined the maximum 
$R_{\rm ms}$ position and  
$r_{\rm X}$ at the very
early epoch of the numerical simulations.   
Figure 9 shows 
$R_{\rm ms}$ as a function of $r/r_{\rm S}$ along the equatorial plane 
at $t = 1 \times 
10^{-4} \tau_{\rm S}$ in the case of $\eta = 0.001r_{\rm S}$.
We can see that $R_{\rm ms}$ has its maximum value
at $r \sim 1.7 r_{\rm S}$.
The position of $r_{\rm X}$
at this time is calculated to be $\sim 1.51 r_{\rm S}$
from the same analysis as shown in Figure 3. Thus
the maximum $R_{\rm ms}$ position and  
$r_{\rm X}$ 
actually locate at almost the same position, which supports the above 
discussion.

\begin{figure} % [H]
\begin{center}
\includegraphics[width=8cm]{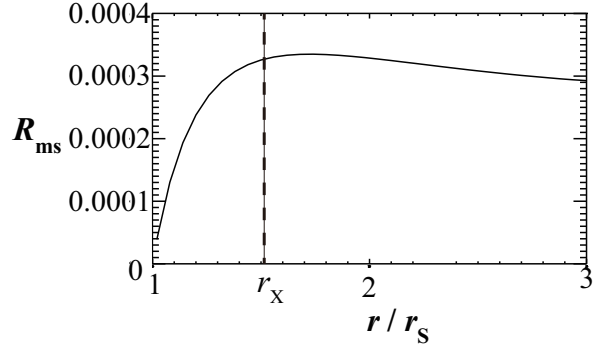}
\caption{
Diffusive slip-through rate of magnetic field lines across plasma
$R_{\rm ms}$ as a function of $r/r_{\rm S}$ along the equatorial plane
at the beginning of the simulation
$t = 1 \times 10^{-4} \tau_{\rm S}$ in the case of
$\eta = 0.001r_{\rm S}$.
The position of the reconnection point $r_{\rm X}$
at this time is calculated to be $\sim 1.51 r_{\rm S}$
from the same analysis as shown in Figure 3.
%$R_{\rm ms}$ value at $r_{\rm X}$ is the reconnection rate, $R_{\rm mr}$.
On the other hand,
$R_{\rm ms}$ has its maximum value
at $r \sim 1.7 r_{\rm S}$.
The maximum $R_{\rm ms}$ position and
$r_{\rm X}$
locate at almost the same position.
\label{fig9}}
\end{center}
\end{figure}

As the stationary models give the magnetic
reconnection rate in the stationary state,
let us calculate this rate and compare our simulation results. Our simulation
results cover 
resistivity $\eta$ values in the range from $1 \times 10^{-5} r_{\rm S}
$ to $0.005r_{\rm S}$, which
corresponds to the global magnetic Reynolds number $S_{\rm e}
\sim 10^5$ to 200.
In the Sweet--Parker mechanism, $S_{\rm e} = 10^5$ yields  
$R_{\rm mr} = 1 / \sqrt{S_{\rm e}} \sim 3.2 \times 10^{-3}$.
In the Petschek model, $S_e = 10^5$ yields 
$R_{\rm mr}$(max) $= \pi / (8 {\rm ln} S_{\rm e}) \sim 3.4 \times 10^{-2}$.
As shown in Figures 4 and 5, $R_{\rm mr}$ is smaller than 0.005 for 
$\eta \leq 1 \times
10^{-4}$ or $S_{\rm e} \geq 10^4$. Thus these magnetic reconnections   
can be regarded as the nonlinear 
growth stage before it settles down to the stationary stage.

With the resistive GRMHD numerical calculations,
drastic phenomena around rapidly spinning 
black holes related with magnetic reconnection
will be revealed.
One of such phenomena is energy extraction from a black hole through
magnetic reconnection (Koide \& Arai 2008).
Magnetic reconnection is expected to occur frequently
around a rapidly rotating
black hole. In this paper, we assumed a Schwarzschild black 
hole, so we will extend our simulation for the case of a Kerr black hole
in the near future. 
We will also perform the longer-term simulations of magnetic reconnection
for larger $S$   
and see if the magnetic reconnection settles down to the
stationary stage from the nonlinear transition stage.
%Performing simulations of
%magnetic reconnection around a black hole
%for larger $S$ (smaller $\eta$) value for longer time,
%and for different settings (different magnetic field strength or magnetic field
%configurations) is also our future work.
Magnetic islands which appear at the later stage
are  
interesting phenomena.
Further studies of them are necessary.

\begin{acknowledgments}
The authors are grateful to the anonymous referee for improving the original
manuscript.   
\end{acknowledgments}

\end{document}